\documentclass[reprint,superscriptaddress,nofootinbib,amsmath,amssymb,floatfix,float,aps,prd]{revtex4-2}
\usepackage{graphicx}
\usepackage{dcolumn}
\usepackage{bm}
\usepackage[colorlinks=true, allcolors=blue]{hyperref}
\usepackage{cleveref}
\usepackage{xcolor}
\usepackage[normalem]{ulem}

\begin{document}

\title[]{{Gravitational waveforms from inspiraling compact binaries in quadratic gravity and their parameterized post-Einstein characterization}}

\author{Matheus F. S. Alves}
\email{matheus.s.alves@edu.ufes.br}
\affiliation{Departamento de Física \& Núcleo de Astrofísica e Cosmologia (Cosmo-Ufes), Universidade Federal do Espírito Santo, Vitória, ES,  29075-910, Brazil.}
\author{L.G. Medeiros}
\email{leo.medeiros@ufrn.br}
\affiliation{Escola de Ci\^encia e Tecnologia, Universidade Federal do Rio Grande
do Norte, Campus Universit\'ario, s/n\textendash Lagoa Nova, CEP 59078-970,
Natal, Rio Grande do Norte, Brazil.}
\author{Davi C. Rodrigues}
 \email{davi.rodrigues@ufes.br}
\affiliation{Departamento de Física \& Núcleo de Astrofísica e Cosmologia (Cosmo-Ufes), Universidade Federal do Espírito Santo, Vitória, ES,  29075-910, Brazil.}
\affiliation{Centro Brasileiro de Pesquisas Físicas (CBPF), R. Xavier Sigaud 150, 22290-180, Rio de Janeiro, RJ, Brazil}
\begin{abstract} 
    We investigate gravitational waveforms from the inspiral phase of compact binary systems within the framework of quadratic gravity and map their deviations from general relativity into the parameterized post-Einstein (ppE) formalism to constrain the theory’s parameters. Quadratic gravity generically includes a massive spin-2 ghost, which leads to ill-defined energy and angular momentum fluxes. Following earlier proposals, we remove these unphysical features by imposing a constraint on the massive spin-2 mode, restricting it to propagate only the same polarizations of general relativity. Within the quadrupole approximation, we derive the radiative degrees of freedom, including massless and massive tensor modes, as well as a massive scalar field. Using the stationary phase approximation, we compute the Fourier-domain waveform of the massless tensor modes and extract the phase corrections. For small deviations from general relativity, we show that both the scalar and massive tensor modes can be consistently embedded into the ppE framework, extending previous results that considered only scalar fields. We derive updated constraints on the parameters of quadratic gravity, finding bounds improved by several orders of magnitude compared to existing limits. Finally, we present forecasts for the sensitivity of the Einstein Telescope to these deviations.

\end{abstract}

\maketitle

\section{INTRODUCTION}
The detection of gravitational waves (GWs)
has opened a unique window to probe gravity in strong-field \cite{LIGOScientific:2016aoc}. From the first observed event in 2015, a binary
black hole merger, more than 100 events have been cataloged so far, including the
detection of a neutron star merger in 2017, with an associated
electromagnetic counterpart, GW170817 \cite{LIGOScientific:2017vwq, Goldstein:2017mmi}, and also a possible NS-BH coalescence event
GW190426, albeit its highest false alarm rate \cite{
LIGOScientific:2020aai, 10.1093/mnras/stab2716, Li:2020pzv, Niu:2021nic}.
These observations have provided robust tests of general relativity under
extreme conditions \cite{Berti:2015itd, Barack:2018yly, Berti:2018cxi,
Yunes:2024lzm}.
Although General Relativity (GR) is the most robust theory of gravity so far, it nonetheless
faces  significant challenges 
from a theoretical standpoint. Theoretically, GR cannot be renormalized
using conventional methods,
it contains singularities \cite{Penrose:1964wq}, and it cannot self-consistently
describe the very early inflationary universe without the inclusion of extra
degrees of freedom \cite{STAROBINSKY198099, Martin:2013tda,
Cuzinatto:2018vjt, Rodrigues-da-Silva:2022qiq}. Moreover, the need to invoke dark matter, dark energy, and, more recently, the Hubble tension \cite{DiValentino:2021izs} has prompted questions as to whether these phenomena signal a possible breakdown in the validity of GR on cosmological scales. These open issues, even within the classical regime, have motivated ongoing efforts to explore extensions and modifications of GR (e.g., for reviews, \cite{1107453984, Capozziello:2017epp, Shankaranarayanan:2022wbx}; and some recent developments can be find in Refs.~\cite{Cuzinatto:2018chu, Bertini:2024onw, Hipolito-Ricaldi:2024xlb, Alves:2025zkk}). 

Higher-order extensions of General Relativity (GR) can be broadly divided into two classes. The first comprises purely metric theories, in which all additional dynamics stem solely from higher powers or derivatives of curvature invariants—these are commonly referred to as higher-order gravity theories. The second class includes frameworks where curvature invariants are coupled to extra dynamical fields, typically scalars, thereby introducing additional degrees of freedom. Prominent examples of the latter are dynamical Chern–Simons (dCS) \cite{Alexander:2009tp} and scalar Gauss–Bonnet (sGB) \cite{Boulware:1986dr} theories, which respectively couple an axion and a dilaton field to parity-violating and parity-conserving curvature invariants. Their well-defined perturbative structure and connections to string-inspired effective actions make them particularly suitable for gravitational-wave tests of GR \cite{Silva:2017uqg, Cano:2021rey, Fernandes:2022zrq, Daniel:2024lev}.

Focusing on the purely metric sector, the theory considered in this work—quadratic gravity—belongs to the first class: it is a higher-order extension of GR that does not introduce additional fields beyond the metric. Within the framework of the Effective Field Theory (EFT) approach to gravity, such theories naturally arise from curvature-squared and higher-derivative operators that encode low-energy remnants of ultraviolet completions \cite{Cano:2023jbk, Liu:2024atc}. While sharing structural similarities with dCS and sGB—particularly regarding the presence of higher-curvature corrections—the model studied here represents the minimal purely geometric extension relevant in strong-field or high-curvature regimes \cite{Cayuso:2023xbc, Endlich:2017tqa}.

Building upon this framework, we now focus on quadratic gravity \cite{Stelle:1976gc, Donoghue:2021cza, Shapiro:2022ugk}, which provides a consistent setting for the renormalization of quantum fields in curved spacetime, albeit at the cost of introducing a ghost degree of freedom. Several strategies have been proposed to address this issue (see \cite{Shapiro:2022ugk, Basile:2024oms} for a comprehensive review), and while a complete resolution remains elusive, promising approaches continue to emerge \cite{Camanho:2014apa, Edelstein:2021jyu}. Simultaneously, quadratic gravity and its extensions (e.g., \cite{deOSalles:2018eon, Rachwal:2021bgb, Asorey:2024oxw}) have been explored as effective field theories that encode high-energy corrections to GR, although alternative viewpoints and limitations have also been discussed \cite{Woodard:2023tgb}.

It is well known that quadratic gravity, at linear order, can be in general decomposed into two tensorial spin-2 modes, massive and massless, and a scalar spin-0 mode \cite{Stelle:1976gc, Stelle:1977ry} (see also \cite{Bueno:2016ypa, Alves:2025zkk} for more recent approaches). Gravitational wave generation from binary black holes in non-relativistic circular orbits was analyzed by some of us in Ref.~\cite{Alves:2022yea}, where constraints on model parameters were derived from observational data, assuming negligible longitudinal tensorial polarizations and ignoring angular momentum dissipation. In Ref.~\cite{Alves:2024gwi} we addressed these limitations by investigating the angular momentum loss process and the influence of longitudinal tensorial modes. The main result of this study is the demonstration that, within the quadrupole approximation, only the longitudinal modes of the massive tensorial term exhibit pathologies, such as negative power emission. By focusing exclusively on the transverse-traceless modes, the study established a self-consistent model that ensures positive energy and angular momentum for the emitted gravitational waves.   

Tests of GR with gravitational wave data are usually divided into consistency tests and parameterized tests \cite{LIGOScientific:2021sio}. Consistency tests, such as residual analyses and inspiral-merger-ringdown checks, are designed to verify whether the observed signals are consistent with the GR predictions. On the other hand, the parameterized tests introduce a deviation into the waveform model to search for possible non-GR effects. A most popular framework for such tests is the parameterized post-Einsteinian (ppE) \cite{Yunes:2009ke, Sampson:2013lpa, Chatziioannou:2012rf, Cornish:2011ys}, which deforms the post-Newtonian (PN) expansion of the waveform phase. The original ppE framework introduces two parameters: the ppE index, characterizing the PN order at which the deformation takes place, and the ppE coefficient, characterizing the magnitude of the deformation \cite{Yunes:2016jcc, Yunes:2009ke, Tahura:2018zuq, Perkins:2020tra}. 

In this paper, we compute ppE waveforms in quadratic gravity and obtain constraints on the theory's parameters. We do this by computing the GW waveforms produced during the inspiral of a compact binary in non-relativistic quasicircular orbits. More precisely, we compute the waveform expressions for the relevant polarizations in the time domain as obtained in \cite{Alves:2022yea}. Considering only the leading PN corrections to the massless tensorial modes, we employ the original ppE framework \cite{Yunes:2009ke}. We calculate the waveforms in the frequency domain using Fourier transforms and the stationary phase method (SPA). 
We show that, by constraining the massive mode to have only $\times$ and $+$ polarizations, the massless tensorial polarizations can be successfully embedded into the ppE framework. These results extend the results obtained in \cite{Liu:2020moh}, where it was possible to map the scalar field to ppE only for sufficiently small mass. In this work, we employed the ppE parameter constraints on Quadratic Gravity to significantly tighten the bounds previously obtained by some of us in \cite{Alves:2022yea}, improving them by several orders of magnitude.

The paper is organized as follows: In Sec. \ref{sec:quadraticgravity}, we describe and obtain the field equations of quadratic gravity and linearize them. In Sec. \ref{sec:waveforms}, we obtain the time-domain gravitational waveforms corresponding to tensorial massive and massless fields as well as the scalar mode. Furthermore, we study the energy loss induced by the emission of gravitational waves and obtain the balance equation. With this, we get the waveforms for the massless tensorial mode in the frequency domain using the Fourier transform and the stationary phase method. In Sec. \ref{sec:PPE}, we obtain post-Einstein parameters in quadratic gravity and find possible constraints on the theory's parameters. At last, in Sec. \ref{sec:conclusion}, we present our final comments.

We use the metric signature $(- + + +)$.

\section{QUADRATIC GRAVITY} \label{sec:quadraticgravity}
From a purely theoretical point of view, GR can be constructed by considering that gravity is described by a $4$-dimensional Riemannian metric theory,
containing the metric $g_{\mu\nu}$ as the only fundamental field, invariant
under diffeomorphisms and whose field equations are of second order (Lovelock
theorem) ~\cite{Clifton:2011jh}. Modifications to GR are achieved by violating any of these
hypotheses. For example, if we violate the hypothesis that the metric is the
only fundamental field and include an extra scalar degree of freedom, we
obtain the Horndeski theories ~\cite{Horndeski:1974wa}. On the other hand, if we consider a
Riemann-Cartan geometry containing an affine connection with a non-vanishing
antisymmetric part, we include torsion in the description of gravity and
obtain the Einstein-Cartan theories ~\cite{Trautman:2006fp}. Alternatively, by allowing the field equations of the theory to be different from second order while preserving all other hypotheses, we find the higher-order gravity theories ~\cite{Cuzinatto:2006hb,Cuzinatto:2016ehv}.

Higher-order gravity theories are characterized by the inclusion of corrections
terms in the Einstein-Hilbert (EH) action that generate higher-order field
equations. Such corrections can be classified according to their
mass/energy scale. Following this classification, the zero-order terms,
mass squared, are given by $R$ and $\Lambda$, and are associated with the
EH action itself with cosmological constant. All first-order corrections,
involving mass to the fourth power terms, are constituted by the invariants
\begin{equation*}
R^{2}\text{,  }R_{\mu\nu}R^{\mu\nu},\text{ }R_{\mu\nu\alpha\beta}R^{\mu
\nu\alpha\beta}\ \text{and }\square R.\label{ordem um}%
\end{equation*}
However, among these four invariants only $2$ contribute to the field
equations. In fact, the term $\square R$ is an explicit surface term, while
the terms $R_{\mu\nu}R^{\mu\nu}$ and $R_{\mu\nu\alpha\beta}R^{\mu\nu
\alpha\beta}$ can be written in terms of the Gauss-Bonnet $G^{2}$ and Weyl
$C^{2}$ invariants:
\begin{align*}
G^{2} &  \equiv R^{2}-4R_{\mu\nu}R^{\mu\nu}+R_{\mu\nu\alpha\beta}R^{\mu
\nu\alpha\beta},\\
C^{2} &  \equiv C_{\mu\nu\alpha\beta}C^{\mu\nu\alpha\beta}=\frac{1}{3}%
R^{2}-2R_{\mu\nu}R^{\mu\nu}+R_{\mu\nu\alpha\beta}R^{\mu\nu\alpha\beta}.%
\end{align*}
Thus,%
\begin{align*}
R_{\mu\nu}R^{\mu\nu}  & =\frac{1}{3}R^{2}-\frac{1}{2}G^{2}+\frac{1}{2}C^{2},\\
R_{\mu\nu\alpha\beta}R^{\mu\nu\alpha\beta}  & =\frac{1}{3}R^{2}-G^{2}+2C^{2}.
\end{align*}
Additionally, in $4$ dimensions the term $G^{2}$ is a topological
invariant and does not contribute to the field equations. Therefore, in the
context of higher-order gravity, the general action that takes into account
all first-order corrections to GR is given by\footnote{We are neglecting the influence of the cosmological constant.}
\begin{eqnarray}
S=-\frac{1}{2\kappa}\int d^{4}x\sqrt{-g}\left[  R+\frac{1}{2}\gamma
R^{2}-\frac{1}{2}\alpha C^2\right]+ S_m. \label{eq:QG_action}%
\end{eqnarray}
This theory is known as quadratic gravity. Here,  $S_{m}$ is the action of the matter fields, $\kappa={8\pi G / c^{4}}$, and $\alpha$,
$\gamma$ are parameters with dimensions of squared length. The field equation
is obtained from (\ref{eq:QG_action}) by varying it with respect to $g^{\mu
\nu}$ \cite{Alves:2022yea}:%
\begin{gather}
R_{\mu \nu }-\frac{1}{2}g_{\mu \nu }R -2\alpha \left[ \nabla ^{\rho }\nabla ^{\beta }C_{\mu \rho \nu \beta }+\frac{1%
}{2}R^{\rho \beta }C_{\mu \rho \nu \beta }\right]
\\ +\gamma \left[ R\left( R_{\mu \nu }-%
\frac{1}{4}Rg_{\mu \nu }\right) +g_{\mu \nu }\nabla _{\rho }\nabla ^{\rho
}R-\nabla _{\mu }\nabla _{\nu }R\right] \nonumber   
 =\kappa T_{\mu \nu }. 
\label{eq:QG_field_equations}
\end{gather}
To obtain the linearized equations on a flat background, we start by writing the metric tensor as usual
\begin{equation}
g_{\mu\nu}= \eta_{\mu\nu} + h_{\mu \nu} = \eta_{\mu\nu}+\left(  \bar{h}_{\mu\nu}-\frac{1}{2}\eta_{\mu\nu}%
\bar{h}\right)  ,\label{eq:usal_decomp}%
\end{equation}
where $\left\vert \bar{h}_{\mu\nu}\right\vert \ll1$ and $\bar{h}=\eta^{\mu\nu
}\bar h_{\mu\nu}$. Using the same approach of Ref.~\cite{Alves:2024gwi, Alves:2025zkk}, we can
decompose $\bar{h}_{\mu\nu}$ in the form 
\begin{equation}
\bar{h}_{\mu\nu}=\tilde{h}_{\mu\nu}+\Psi_{\mu\nu}-\eta_{\mu\nu}\left(
\Phi+\frac{1}{6}\Psi\right) \, \label{eq:decomp},
\end{equation}
where $\Psi=\eta^{\mu\nu}\Psi_{\mu\nu}$.
Besides, assuming the Teyssandier gauge \cite{Teyssandier1989}, considering vaccum solution, and taking into account the residual gauge freedom, we can impose the conditions\footnote{{As we show in \cite{Alves:2025zkk}, there is a different gauge for which $\partial^\mu \Psi_{\mu \nu} = \Psi = 0$ always, even inside matter. However, $\Psi_{\mu \nu}$ in this case is subjected to a different effective energy-momentum tensor that depends on the operator $\Box^{-1}$. In this work, we focus on the Teyssandier gauge.}}
 \cite{Alves:2025zkk}
\begin{equation}
\partial^{\mu}\tilde{h}_{\mu\nu}=\partial^{\mu}\Psi_{\mu\nu}=0\text{ and
}\tilde{h}=\Psi=0.
\end{equation}
Under these assumptions, the wave equations for the
fields $\tilde{h}_{\mu\nu}$, $\Psi_{\mu\nu}$, $\Phi$ are given by
\begin{eqnarray}
\square\tilde{h}_{\mu\nu}=-2\kappa T_{\mu\nu},\label{eq:eq_h}
\\
\left(  \square-m_{\Psi}^{2}\right)  \Psi_{\mu\nu}=2\kappa T_{\mu\nu},\text{
}\label{eq:eq_psi}\\
\left(  \square-m_{\Phi}^{2}\right)  \Phi=-\frac{1}{3}\kappa T,\text{
}\label{eq:eq_phi}
\end{eqnarray}
with $m_{\Psi}^{2}=\frac{1}{\alpha}$ and $m_{\Phi}^{2}=\frac{1}{3\gamma}$.
\ The above equations characterize $\Phi$, $\tilde{h}_{\mu\nu}$ and
$\Psi_{\mu\nu}$ as a massive spin-0 field, a massless spin-2 field, and a
massive spin-2 field, respectively. Notice that the massive degrees of freedom vanish in the limit $\alpha$, $\gamma$ $\rightarrow0$; in which case we recover results from General Relativity (GR).

\section{Gravitational waveforms in quadratic gravity} \label{sec:waveforms}

\subsection{Time-domain GW waveforms}

This section aims to calculate gravitational waveforms in QG up to the lowest post-Newtonian order. For this goal, we study the emission of GWs produced by the inspiral phase of a binary black hole system in the
approximations of point masses, circular orbit, and non-relativistic dynamics.
Such systems lose energy due to gravitational radiation. Previously, this process was described in detail for QG in \ Ref.~\cite{Alves:2022yea}.
Using the method of Green's function and multipole expansion, we obtain the solutions of the field equations (\ref{eq:eq_h}), (\ref{eq:eq_psi}) and (\ref{eq:eq_phi}) in a region devoid of sources and the quadrupole approximation as \cite{Polyanin2002, scharf2014finite, Alves:2022yea, Alves:2024gwi} 
\begin{align}
\tilde{h}_{ij}^{TT}\left(  \mathbf{x},t\right)   &  =\frac{1}{r}\frac{\kappa
}{4\pi}\Lambda_{ij,kl}\left(  \mathbf{\hat{n}}\right)  \ddot{M}^{kl}\left(
t_{r}\right)  ,\label{spin2 massless}\\
\Psi_{ij}\left(  \mathbf{x},t\right)   &  =-\frac{1}{r}\frac{\kappa}{4\pi
}\ddot{M}_{ij}\left(  t_{r}\right)  \label{spin2massive}\\
&  +\frac{\kappa}{4\pi}m_{\Psi}\int_{0}^{\infty}d\bar{t}_{r}F_{\Psi}\left(
\bar{t}_{r}\right)  \ddot{M}_{ij}\left(  \zeta\right)  ,\nonumber\\
\Phi^{Q}\left(  \mathbf{x},t\right)   &  =\frac{\kappa}{24\pi r}n_{i}%
n_{j}\mathcal{\ddot{M}}^{ij}\left(  t_{r}\right)  \label{spin0}\\
&  -\frac{\kappa}{24\pi}m_{\Phi}\int_{0}^{\infty}d\bar{t}_{r}F_{\Phi}\left(
\bar{t}_{r}\right)  \mathcal{\ddot{M}}_{ij}\left(  \zeta\right)  ,\nonumber
\end{align}
where $\Lambda_{ij,kl}$ is the lambda projection tensor in TT gauge
\cite{Maggiore2007}, $t_{r}=t-\frac{r}{c}$ is the retarded time,
$\zeta\equiv t_{r}-\bar{t}_{r}$, $n_{i}$ points in the direction of
propagation of the wave, and
\begin{equation}
F_{X}\left(  \bar{t}_{r}\right)  =\frac{J_{1}\left(  m_{X}c\sqrt{2\bar{t}_{r}%
}\sqrt{\frac{\bar{t}_{r}}{2}+\frac{r}{c}}\right)  }{\sqrt{2\bar{t}_{r}}%
\sqrt{\frac{\bar{t}_{r}}{2}+\frac{r}{c}}}\label{F}%
\end{equation}
with $J_1$ representing the Bessel function of the first kind.
Furthermore, the quadrupole moment for the tensorial modes are the usual from
GR \cite{Maggiore2007}%
\begin{equation}
M^{ij}=\frac{1}{c^{2}}\int d^{3}xT^{00}\left(  t,\mathbf{x}\right)  x^{i}%
x^{j},\label{eq:quad_momentum_tensor}%
\end{equation}
and the quadrupole moment for the scalar mode is built with the trace of the
energy-momentum tensor as \cite{Alves:2022yea, Vilhena:2021bsx}
\begin{equation}
\mathcal{M}^{ij}\left(  t\right)  \equiv\frac{1}{c^{2}}\int d^{3}%
\mathbf{x}^{\prime}T\left(  \mathbf{x}^{\prime},t\right)  x^{\prime
i}x^{\prime j}.\label{eq:quad_momentum_scalar}%
\end{equation}

As discussed in Ref.~\cite{Alves:2024gwi} and well-established in the literature~\cite{Stelle:1976gc}, the presence of the term \( C_{\mu\nu\alpha\beta}C^{\mu\nu\alpha\beta} \) in quadratic gravity (QG) introduces an Ostrogradsky instability, leading to a non--positive-definite energy. In the context of gravitational-wave emission, this manifests as a negative radiated power~\cite{Alves:2022yea, Alves:2024gwi}, signaling a clear physical inconsistency.

The treatment of this instability was analyzed in detail in Ref.~\cite{Alves:2024gwi} (Sec.~III.D). In the quadrupole approximation, the instability originates from the longitudinal propagation modes of the massive spin-2 field \(\Psi_{ij}\). A consistent way to remove this pathology is to eliminate such modes by projecting \(\Psi_{ij}\) onto its transverse-traceless (TT) subspace through the operator \(\Lambda_{ij,kl}(\mathbf{\hat{n}})\):\footnote{The gauge structure of the massive spin-2 sector allows one to take \(\Psi_{ij}\) as traceless.}

\begin{equation}
\Psi_{ij} = \Psi_{ij}^{L} + \Psi_{ij}^{TT}
\;\longrightarrow\;
\Lambda_{ij,kl}\!\left(\mathbf{\hat{n}}\right)
\left(\Psi_{kl}^{L} + \Psi_{kl}^{TT}\right)
= \Psi_{ij}^{TT}.
\label{eq:QC_solve_inst}
\end{equation}

Although this procedure is mathematically identical to that applied to the massless spin-2 field \(\tilde{h}_{ij}\), its physical motivation is distinct. For \(\tilde{h}_{ij}\), only the TT components are retained due to the residual gauge freedom inherent to the massless sector. In contrast, for the massive tensor \(\Psi_{ij}\), the elimination of the longitudinal components \(\Psi_{ij}^{L}\) is imposed explicitly to remove unphysical (ghostlike) modes responsible for the Ostrogradsky instability. This restriction could, in principle, be introduced at the level of the action, but for simplicity it is implemented directly in the field equations.

Consequently, although QG originally propagates six degrees of freedom---five associated with the tensorial fields \(\tilde{h}_{ij}^{TT}\) and \(\Psi_{ij}\), and one scalar---the imposition of the TT projection reduces the spectrum to three physical modes: two tensorial (\(\tilde{h}_{ij}^{TT}\) and \(\Psi_{ij}^{TT}\)) and one scalar. This procedure effectively removes the pathological contributions, ensuring a stable and energetically consistent gravitational-wave sector.

The stress-energy tensor for a nonrelativistic binary point-mass system
$m_{A}$ in the center-of-mass frame can be expressed as%
\begin{equation}
T_{\mu\nu}=\sum_{A=1}^{2}m_{A}c^{2}\delta_{\mu}^{0}\delta_{\nu}^{0}
\delta^{\left(  3\right)  }\left(  \mathbf{x}-\mathbf{x}_{A}\left(  t\right)
\right)  , \label{eq:energy_mon_tensor}
\end{equation}
where $\mathbf{x}_{A}\left(  t\right)$ is the vector representing the
trajectory of particle $A$. From this energy-momentum tensor and its trace, we
can obtain the mass moments for scalar and tensorial modes in the
center-of-mass frame. So, by equations (\ref{eq:quad_momentum_tensor}) and
(\ref{eq:quad_momentum_scalar}), we get
\begin{equation}
\mathcal{M}^{ij}\left(  t\right)  =-\mu x_{0}^{i}\left(  t\right)  x_{0}%
^{j}\left(  t\right)  =-M^{ij}\left(  t\right)  , \label{eq:mass_mon_binary}%
\end{equation}
where $m=m_{1}+m_{2}$ is the total mass, $\mu=\frac{m_{1}m_{2}}{m}$ is the
reduced mass, and $x_{0}^{i}\left(  t\right)  $ is the relative coordinate
$\mathbf{x}_{0}=\mathbf{x}_{1}-\mathbf{x}_{2}$.

To obtain the waveforms, we need to determine the trajectory
$x_{0}^{i}\left(  t\right)  $. \ We will consider a circular orbit of radius
$R$ and angular frequency $\omega_{s}$ positioned along the $xy$-plane.
Accordingly, the only non-zero mass moments are $M_{11}=-M_{22}$ and
$M_{12}=M_{21}$ \cite{Alves:2022yea, Vilhena:2021bsx}. The spin-$0$,
massive spin-$2$, and massless spin-$2$ modes are obtained by substituting the
non-null mass moments in Eqs. (\ref{spin0}), (\ref{spin2massive}),
(\ref{spin2 massless}), and then calculating the integrals that contain the
functions $F_{X}$ function (\ref{F}). Carrying out all these calculations,
considering the chirp mass $\mathcal{M}_{c}=m^{\frac{2}{5}}\mu^{\frac{3}{5}}$ and the
Kepler's third law $\omega_{s}^{2}=\frac{Gm}{R^{3}}$, we obtain  
\begin{widetext}
\begin{eqnarray}
\tilde{h}_{+}\left(  \mathbf{x},t\right)&=\frac{4c}{r}\left(
\frac{G\mathcal{M}_{c}}{c^{3}}\right)  ^{\frac{5}{3}}\omega_{s}^{2/3}\left(
\frac{1+\cos^{2}\theta}{2}\right)  \cos\left[  2\omega_{s}\left(  t-\frac
{r}{c}\right)  +2\phi\right]  ,\label{h plus}\\
\tilde{h}_{\times}\left(  \mathbf{x},t\right)&=\frac{4c}{r}\left(
\frac{G\mathcal{M}_{c}}{c^{3}}\right)  ^{\frac{5}{3}}\omega_{s}^{2/3}\cos\theta
\sin\left[  2\omega_{s}\left(  t-\frac{r}{c}\right)  +2\phi\right]  ,
\label{h cross}%
\end{eqnarray}
for massless tensorial modes,%
\begin{align}
\Psi_{+}\left(  \mathbf{x},t\right)   &  =\left\{
\begin{array}
[c]{c}%
-\frac{4c}{r}\left(  \frac{G\mathcal{M}_{c}}{c^{3}}\right)  ^{\frac{5}{3}}\omega
_{s}^{2/3}\left(  \frac{1+\cos^{2}\theta}{2}\right)  \cos\left(  2\omega
_{s}t+2\phi\right)  \exp\left[  -m_{\Psi}r\sqrt{1-\left(  \frac{2\omega_{s}
}{m_{\Psi}c}\right)  ^{2}}\right]  ,\text{ \ \ }2\omega_{s}<m_{\Psi}c\\
-\frac{4c}{r}\left(  \frac{G\mathcal{M}_{c}}{c^{3}}\right)  ^{\frac{5}{3}}\omega
_{s}^{2/3}\left(  \frac{1+\cos^{2}\theta}{2}\right)  \cos\left[  2\omega
_{s}\left(  t-\left(  \frac{r}{c}\right)  \sqrt{1-\left(  \frac{m_{\Psi}
c}{2\omega_{s}}\right)  ^{2}}\right)  +2\phi\right]  ,\text{ \ \ \ }
2\omega_{s}>m_{\Psi}c
\end{array}
\right.  ,\label{Psi plus}\\
\Psi_{\times}\left(  \mathbf{x},t\right)   &  =\left\{
\begin{array}
[c]{c}%
-\frac{4c}{r}\left(  \frac{G\mathcal{M}_{c}}{c^{3}}\right)  ^{\frac{5}{3}}\omega
_{s}^{2/3}\cos\theta\sin\left(  2\omega_{s}t+2\phi\right)  \exp\left[
-m_{\Psi}r\sqrt{1-\left(  \frac{2\omega_{s}}{m_{\Psi}c}\right)  ^{2}}\right]
,\text{ \ \ }2\omega_{s}<m_{\Psi}c\\
-\frac{4c}{r}\left(  \frac{G\mathcal{M}_{c}}{c^{3}}\right)  ^{\frac{5}{3}}\omega
_{s}^{2/3}\cos\theta\sin\left[  2\omega_{s}\left(  t-\left(  \frac{r}%
{c}\right)  \sqrt{1-\left(  \frac{m_{\Psi}c}{2\omega_{s}}\right)  ^{2}%
}\right)  +2\phi\right]  ,\text{ \ \ \ }2\omega_{s}>m_{\Psi}c
\end{array}
\right.  , \label{Psi cross}%
\end{align}
for the massive tensorial modes and
\begin{equation}
\Phi\left(  \mathbf{x},t\right)  =\left\{
\begin{array}
[c]{c}%
\frac{2c}{3r}\left(  \frac{G\mathcal{M}_{c}}{c^{3}}\right)  ^{\frac{5}{3}}\omega
_{s}^{2/3}\sin^{2}\theta\cos\left(  2\omega_{s}t+2\phi\right)  \exp\left[
-m_{\Phi}r\sqrt{1-\left(  \frac{2\omega_{s}}{m_{\Phi}c}\right)  ^{2}}\right]
,\text{ \ \ }2\omega_{s}<m_{\Phi}c\\
\frac{2c}{3r}\left(  \frac{G\mathcal{M}_{c}}{c^{3}}\right)  ^{\frac{5}{3}}\omega
_{s}^{2/3}\sin^{2}\theta\cos\left[  2\omega_{s}\left(  t-\left(  \frac{r}%
{c}\right)  \sqrt{1-\left(  \frac{m_{\Phi}c}{2\omega_{s}}\right)  ^{2}%
}\right)  +2\phi\right]  ,\text{ \ \ \ }2\omega_{s}>m_{\Phi}c
\end{array}
\right.  , \label{Phi}
\end{equation}
for massive scalar mode.
\end{widetext}

As discussed in Ref.\cite{Alves:2022yea}, the most relevant point of the
solutions above is the existence of two distinct regimes for massive modes.
The first regime is a damping one, which occurs when $2\omega_{s}<m_{X}c$,
and the massive modes only contribute with a temporal modulation. The second regime, called oscillatory regime, occurs when $2\omega_{s}>m_{X}c$, and the
massive modes emit gravitational waves. Furthermore, note that in the
oscillatory regime, the solutions $h_{+,\times}$\ and $\Psi_{+,\times}$\ are
waves with the same amplitude and frequency but with different wave numbers.
This difference enables an interpretation of destructive interference effects, and in the limit as $m_{\Psi}\rightarrow0$, this destructive
interference becomes complete, leading to the absence of tensorial mode emission.

\subsection{Evolution of binary systems}

Following Refs.~\cite{Saffer:2017ywl, Isaacson:1968zza}, we derive the gravitational-wave (GW)
radiation power of a binary system in quadratic gravity (QG). The GW energy--momentum tensor is~\cite{Alves:2022yea}
\begin{eqnarray}
t_{\mu \alpha } &=&\frac{c^{4}}{32\pi G}\left\langle \partial _{\mu }\tilde{h%
}_{\nu \beta }\partial _{\alpha }\tilde{h}^{\nu \beta }\right\rangle 
-\frac{c^{4}}{32\pi G}\left\langle \partial _{\mu }\Psi _{\nu \beta}
\partial _{\alpha }\Psi ^{\nu \beta }\right\rangle   \nonumber \\[4pt]
&&+\frac{3c^{4}}{16\pi G}\left\langle \partial _{\mu }\Phi \, \partial _{\alpha}
\Phi \right\rangle ,
\label{eq:energy_mon_gw}
\end{eqnarray}
where $\langle \dots \rangle$ denotes the average over several GW wavelengths.
The radiated power per unit solid angle is then given by
\begin{equation}
\frac{dP}{d\Omega}=-cr^{2}t_{01},
\label{eq:radiated_power}
\end{equation}
with $x^1 = r$ in spherical coordinates. By computing the derivatives in Eq.~(\ref{eq:energy_mon_gw})
and performing the spatial--temporal averages, one obtains~\cite{Alves:2022yea}
\begin{widetext}
\begin{align}
\frac{dP}{d\Omega} &  =-\frac{2G\mu^{2}R^{4}\omega_{s}^{6}}{\pi c^{5}}\!\left\{
\!\left[  \left(  \frac{1+\cos^{2}\theta}{2}\right)^{2}\!+\cos^{2}\theta\right]
\!\!\left(  1-\Theta\!\left(  2\omega_{s}-m_{\Psi}c\right)
\!\sqrt{1-\!\left(\!\frac{m_{\Psi}c}{2\omega_{s}}\!\right)^{2}}\right)
\right. \nonumber \\[4pt]
&  \left.
+\,\Theta\!\left(  2\omega_{s}-m_{\Phi}c\right)
\sqrt{1-\!\left(\!\frac{m_{\Phi}c}{2\omega_{s}}\!\right)^{2}}
\frac{\sin^{4}\theta}{12}\!
\right\},
\label{eq: Pot irr ang}
\end{align}
\end{widetext}
where the Heaviside functions $\Theta(2\omega_{s}-m_{\Psi}c)$ and
$\Theta(2\omega_{s}-m_{\Phi}c)$ indicate that only oscillatory regimes contribute to the radiated power.
After integration over the solid angle, the total emitted power becomes
\begin{eqnarray}
P &=&-\frac{32c^{5}}{5G}\left( \frac{G\mathcal{M}_{c}\omega _{s}}{c^{3}}\right) ^{10/3}
  \label{eq: Pot irr Tot} \\[4pt]
&&+\Theta \left( 2\omega _{s}-m_{\Psi }c\right)
  \frac{32c^{5}}{5G}\left( \frac{G\mathcal{M}_{c}\omega _{s}}{c^{3}}\right) ^{10/3}
  \sqrt{1-\left( \frac{m_{\Psi }c}{2\omega _{s}}\right)^{2}}  \nonumber \\[4pt]
&&-\frac{\Theta \left( 2\omega _{s}-m_{\Phi }c\right)}{18}
  \frac{32c^{5}}{5G}\left( \frac{G\mathcal{M}_{c}\omega _{s}}{c^{3}}\right) ^{10/3}
  \sqrt{1-\left( \frac{m_{\Phi }c}{2\omega _{s}}\right)^{2}}.  \nonumber
\end{eqnarray}

In general, the nonrelativistic orbital dynamics of modified gravity
theories exhibit corrections to the Newtonian potential. For instance, in the
case of quadratic gravity, besides the standard $\frac{1}{r}$ term, the potential may include
two Yukawa-type contributions. However, for black holes (BHs), these modifications are not always present.
A simple illustration is provided by the $R+\gamma R^{2}$ model, where it was shown in Ref.~\cite{Nelson:2010ig} that
the exterior solution of a static, spherically symmetric BH is
necessarily Schwarzschild (\(R_{\mu\nu}=0\)).
This result follows from the boundary conditions imposed by the event horizon,
which enforce a vanishing scalar curvature outside the BH. Consequently, the
nonrelativistic potential in this case reduces to the Newtonian one, with the Yukawa correction
\(e^{-\frac{m_{\Phi}r}{3r}}\) absent.

In the full quadratic theory, the situation is more subtle, since the field equations
can admit exterior geometries distinct from Schwarzschild~\cite{Lu:2015cqa, Lu:2015psa}.
Nonetheless, Ref.~\cite{Lu:2015psa} showed that for sufficiently small values of 
$\alpha$ (the coupling constant associated with the Weyl-squared term), the Schwarzschild solution becomes isolated, meaning that no nearby asymptotically flat alternatives exist.
Hence, even for \(\alpha \neq 0\), one may consistently consider \(R_{\mu\nu}=0\)
as the exterior BH solution. For larger \(\alpha\), additional solutions could exist,
but the conditions for their realization remain undetermined~\cite{Lu:2015psa}.
Accordingly, we adopt the Schwarzschild metric as the static exterior geometry for BH binaries in QG,
resulting in a nonrelativistic orbital dynamics governed solely by the Newtonian potential.

The adoption of the Schwarzschild solution and the resulting absence of Yukawa-type corrections have direct implications for the multipolar expansion. In general, massive scalar modes would produce dipolar radiation, typically entering at $-1$PN order or lower, which could in principle impose stronger constraints on the field masses than those derived from the $0$PN terms. In the present framework, however, such $-1$PN contributions would arise from Yukawa-type modifications to the orbital potential (see, e.g., Ref.~\cite{Kim:2019sqk}), and they are therefore absent. Consequently, dipolar radiation vanishes, and the leading radiative effects arise solely from the quadrupolar terms.

Taking these considerations into account, for a binary system composed of two spherically symmetric black holes, the balance equation can be expressed as
\begin{eqnarray}
\dot{\omega _{s}} &=&\frac{96}{5}\left( \frac{G\mathcal{M}_{c}}{c^{3}}\right) ^{5/3}\omega _{s}^{11/3}  \label{eq: BalanceEq} \\[4pt]
&&-\Theta \left( 2\omega _{s}-m_{\Psi }c\right)
  \frac{96}{5}\left( \frac{G\mathcal{M}_{c}}{c^{3}}\right) ^{5/3}\omega _{s}^{11/3}
  \sqrt{1-\left( \frac{m_{\Psi }c}{2\omega _{s}}\right)^{2}}  \nonumber \\[4pt]
&&+\frac{\Theta \left( 2\omega _{s}-m_{\Phi }c\right)}{18}
  \frac{96}{5}\left( \frac{G\mathcal{M}_{c}}{c^{3}}\right) ^{5/3}\omega _{s}^{11/3}
  \sqrt{1-\left( \frac{m_{\Phi }c}{2\omega _{s}}\right)^{2}}.  \nonumber
\end{eqnarray}

This expression will be central to the derivation of the frequency-domain
gravitational waveforms presented in the next section.

\subsection{Frequency-domain gravitational waveforms}
When analyzing gravitational waveforms in comparison to
observations, it is standard to apply a Fourier transformation of GW waveforms
with a frequency $f$. Furthermore, the time evolution
of the orbital frequency, derived in Eq. (\ref{eq: BalanceEq}), is
required. The Fourier transform can be computed using the stationary phase
approximation (SPA). The SPA is applicable because, during the inspiral, the
change in orbital frequency over a single period is negligible
\cite{Maggiore2007}.

We only consider the Fourier transforms of the massless tensorial
polarizations. This choice is motivated by two main reasons:
first, as shown in Eq. (\ref{eq: Pot irr Tot}), the contribution of the massive
scalar field is significantly smaller compared to that of the spin-$2$
fields.\footnote{{See the factor $\frac{1}{18}$ in the third line of Eq. (\ref{eq: BalanceEq}).}} Moreover, the interference interpretation of the spin-$2$ waves gives
us combinations of the waveforms (\ref{h cross}), (\ref{Psi cross}),
(\ref{h plus}) and (\ref{Psi plus}) that would be detected with the
interference patterns shown in \cite{Alves:2022yea}.
Since these interference patterns were not detected, we will consider that the massive spin-2 waves are generated very close to the merger. Due to their propagation speed being slower than the speed of light, these waves would not have arrived in time to combine with the massless waves and thus remain undetected directly. (more details in Ref.~\cite{Alves:2022yea}). In this context, we focus on the propagation of massless waves and analyze how the massive fields slightly modify them, treating their effect as a perturbative correction to General Relativity, i.e., when $\frac{m_{\Phi,\psi}c}{2\omega_{s}}\sim1$.

Let's proceed with the Fourier transformation:
\begin{equation}
h_{\lambda}\left(  f\right)  =\int\tilde{h}_{\lambda}\left(  t\right)
e^{i2\pi ft}dt,\label{eq: Fourier_h_plus}%
\end{equation}
where $\lambda=+,\times$. For the mode $\tilde{h}_{+}$, using the SPA, we have%
\begin{equation}
\tilde{h}_{+}\left(  f\right)  \simeq A_{+} e^{i\Sigma_{+}},\label{eq:SPA_h_plus}%
\end{equation}
where
\begin{eqnarray*}
A_+ = \frac{2c}{r}\left(  \frac{G\mathcal{M}_{c}}{c^{3}%
}\right)  ^{\frac{5}{3}}\left(  \frac{1+\cos^{2}\theta}{2}\right)  \omega
_{s}^{2/3}\left(  t_{\ast}\right)  \sqrt{\frac{\pi}{\dot{\omega_{s}}\left(
t_{\ast}\right)  }},
\end{eqnarray*}
and $t_{\ast}$ is the phase stationary point determined by
\begin{equation}
\frac{d}{dt}\left[  -2\Phi\left(  t\right)  +2\pi ft\right]  _{t=t_{\ast}%
}=0\Longrightarrow\omega_{s}\left(  t_{\ast}\right)  =\pi f,
\end{equation}
and
\begin{equation}
\Sigma_{+}=2\pi fr-2\Phi\left(  t_{\ast}\right)  +2\pi ft_{\ast}-\frac{\pi}%
{4},\label{eq: GW_phase}%
\end{equation}
where $\Phi\left(  t\right)  =\omega_{s}\left(  t-\frac{r}{c}\right)  $. We use the relation
\begin{equation}
    2\pi t_{\ast}-2\Phi(t_{\ast})=\int_{\infty}^{\pi f } \frac{2\pi f -2\omega}{\dot{\omega}}d\omega+2\pi t_c - 2\Phi_c
\end{equation}
to express $t_\ast$ in terms of $f$, where $t_c$ is determined by $\omega(t_c)=\infty$ and $\Phi_c = \Phi(t_c)$.

Therefore, the phase $\Sigma_{+}$ can be write as:%
\begin{equation}
\Sigma_{+}=2\pi(t_c+\frac{r}{c})- 2\Phi_c -\frac{\pi}{4}+ \int_{\infty}^{\pi f } \frac{2\pi f -2\omega}{\dot{\omega}}d\omega.\label{eq: phase_h_plus_final}%
\end{equation}

Similarly, the Fourier-transformed mode of $\tilde{h}_{\times}$ is given by%
\begin{equation}
\tilde{h}_{\times}\left(  f\right)  \simeq A_\times e^{i\Sigma_{\times}},
\end{equation}
where
\begin{eqnarray*}
  A_\times = \frac{2c}{r}\left(  \frac{G\mathcal{M}_{c}%
}{c^{3}}\right)  ^{\frac{5}{3}}\cos^{2}\theta\omega_{s}^{2/3}\left(  t_{\ast
}\right)  \sqrt{\frac{\pi}{\dot{\omega_{s}}\left(  t_{\ast}\right)  }%
},
\end{eqnarray*}
and
\begin{equation}
\Sigma_{\times}=\Sigma_{+}+\frac{\pi}{2}.
\end{equation}
To evaluate the integral in Eq. (\ref{eq: phase_h_plus_final}), we need to use
the expression for changing the orbital frequency (\ref{eq: BalanceEq}). 

\subsubsection{GR + massive spin-0 mode}
When only the scalar field is in an oscillatory regime, the equation
(\ref{eq: BalanceEq}) is given by%
\begin{equation}
\dot{\omega_{s}}=\frac{96}{5}\left(  \frac{G\mathcal{M}_{c}}{c^{3}}\right)  ^{\frac
{5}{3}}\omega_{s}^{\frac{11}{3}}\left[ 1+\frac{1}{18}\sqrt{1-\left(
\frac{m_{\Phi}c}{2\omega_{s}}\right)  ^{2}}\right]  .
\label{eq: BalanceEQ_phi}%
\end{equation}
For the amplitude, we have%
\begin{equation}
\tilde{h}_{+}\left(  f\right)  \simeq\sqrt{\frac{5\pi}{24}}\frac{c}{r}\left(
\frac{G\mathcal{M}_{c}}{c^{3}}\right)  ^{\frac{5}{6}}\left(  \pi f\right)  ^{-\frac
{7}{6}}\left(  \frac{1+\cos^{2}\theta}{2}\right)  A_{c}e^{i\Sigma_{+}%
},
\end{equation}
where
\begin{equation}
    A_{c}=\left[  1+\frac{1}{18}\sqrt{1-\left(  \frac{m_{\Phi}c}{2\pi f}\right)
^{2}}\right]  ^{-\frac{1}{2}},
\end{equation}
is the correction due to extra terms, and for the phase, substituting Eq.
(\ref{eq: BalanceEQ_phi}) into (\ref{eq: phase_h_plus_final}), we get
\begin{eqnarray}
\Sigma_{+}=2\pi f\left(  t_{c}+\frac{r}{c}\right)  -2\Phi_{c} -\frac{\pi}{4}\label{eq:phase_phi_case_new}\\
+\frac{5}{96}\left(  \frac{G\mathcal{M}_{c}}{c^{3}}\right)  ^{-\frac{5}{3}}I\left(
f\right)  ,\nonumber%
\end{eqnarray}
where
    \begin{equation}
I\left(  f\right)  =\int_{\infty}^{\pi f}\frac{\left(  2\pi f-2\omega_{s}\right)}{\omega_{s}^{\frac{11}{3}}\left[  1+\frac{1}{18}\sqrt{1-\left(  \frac{m_{\Phi
}c}{2\omega_{s}}\right)  ^{2}}\right]}d\omega_{s}%
,\label{eq:integral_phi_case}%
\end{equation}
is the integral that we need to evaluate. For this, we consider that the
scalar field is a small correction to General Relativity, and obtain%
\begin{equation}
I\left(  f\right)  \approx\int_{\infty}^{\pi f}\frac{\left(  2\pi f-2\omega
_{s}\right)}{\omega_{s}^{\frac{11}{3}}}  \left[  1-\frac{1}{18}\sqrt{1-\left(
\frac{m_{\Phi}c}{2\omega_{s}}\right)  ^{2}}\right]  d\omega_{s}%
.\label{eq:integral_phi_app}%
\end{equation}
Defining%
\begin{equation}
x=\left(  \pi f\right)  ^{2},\text{ }m=\left(  \frac{m_{\Phi}c}{2}\right)
^{2},
\end{equation}
and solving integral (\ref{eq:integral_phi_app}) we obtain
\begin{eqnarray}
I\left( x\right)  &=&\frac{x^{-\frac{5}{6}}}{120}\left[ 54-8F_{2}^{\text{ \ }%
1}\left( -\frac{1}{2},\frac{5}{6},\frac{11}{6},\frac{m}{x}\right) \right. 
\label{eq:integral_solved_exp} \\
&&\left. +5F_{2}^{\text{ \ }1}\left( -\frac{1}{2},\frac{4}{3},\frac{7}{3},%
\frac{m}{x}\right) \right] ,  \notag
\end{eqnarray}
where $F_{2}^{\text{ \ }1}\left(  x\right)  $ is the hypergeometric function.
The next step is to expand Eq. (\ref{eq:integral_solved_exp}) around general
relativity and obtain the expression for phase considering the scalar
correction. We can use the point $x_{0}=am$ with $a>1$. In this case, we obtain\footnote{Although an expansion around \( x_0 = m \) can be performed, the resulting expression is a Puiseux series, not a Taylor series, which prevents its inclusion within the ppE framework.} 
\begin{equation}
I\left(  x\right)  =\frac{x^{-\frac{5}{6}}}{120}\left[  C_{0}+C_{1}\left(  \frac
{x}{am}-1\right) +...\right]  ,
\end{equation}
where
\begin{align}
{C}_{0}& =54-8F_{2}^{\text{ \ }1}\left( -\frac{1}{2},\frac{5}{6},\frac{11%
}{6},\frac{1}{a}\right)   \label{eq:C_0} \\
& +5F_{2}^{\text{ \ }1}\left( -\frac{1}{2},\frac{4}{3},\frac{7}{3},\frac{1}{a%
}\right) ,  \nonumber \\
{C}_{1}& =\frac{10}{77a}\left[ 11 F_{2}^{\text{ \ }1}\left( \frac{1}{2},%
\frac{7}{3},\frac{10}{3},\frac{1}{a}\right) \right.   \label{eq:C_1} \\
& \left. - 14F_{2}^{\text{ \ }1}\left( \frac{1}{2},\frac{11}{6},\frac{17}{6},%
\frac{1}{a}\right) \right] .  \nonumber
\end{align}
Note that for any value of $a>1$ we have no fractional exponents, i.e.
$I\left(  x\right)  $ is in fact a Taylor series. This allows us to fit the
solution into the ppE framework. Therefore, for the phase we have
\begin{align}
\Sigma_{+}&=2\pi f\left(  t_{c}+\frac{r}{c}\right)  -2\Phi_{c}-\frac{\pi}
{4}+\frac{3}{128}\left(  \frac{G\mathcal{M}_{c}\pi f}{c^{3}}\right)  ^{-\frac{5}{3}%
}\times \nonumber\\
&\left[  \frac{1}{54}C_{0}+\frac{1}{54}C_{1}\left(  \frac{\left(  2\pi
f\right)  ^{2}}{am_{\Phi}^{2}c^{2}}-1\right)  +...\right], \label{eq:phase_scalar_complete}
\end{align}
and the amplitude, we perform the same expansion, obtaining the expression
\begin{equation}
A_{+}\left(  f\right)  \simeq\sqrt{\frac{5\pi}{24}}\frac{c}{r}\left(
\frac{G\mathcal{M}_{c}}{c^{3}}\right)  ^{\frac{5}{6}}\left(  \pi f\right)  ^{-\frac
{7}{6}}\left(  \frac{1+\cos^{2}\theta}{2}\right)  A_{c},
\end{equation}
where
\begin{equation}
A_{c}\approx1-\frac{1}{18a^{\frac{3}{2}}\sqrt{a-1}}\left(  \frac{a\left(
2a-3\right)  }{4}+\frac{\left(  \pi f\right)  ^{2}}{\left(  m_{\Phi}c\right)
^{2}}\right)  .\label{eq:amplitude_scalar_complete}
\end{equation}
\subsubsection{GR + massive spin-2 mode}
The procedure for the massive tensorial degree of freedom follows closely that of the scalar case discussed in the previous subsection. 
In this case, however, the additional spin-2 field contributes with the opposite sign to the balance equation.
Accordingly, Eq.~(\ref{eq: BalanceEq}) takes the form
\begin{equation}
\dot{\omega}_{s}=\frac{96}{5}
\left(\frac{G\mathcal{M}_{c}}{c^{3}}\right)^{\!\frac{5}{3}}
\omega_{s}^{\frac{11}{3}}
\!\left[1-\sqrt{1-\left(\frac{m_{\Psi}c}{2\omega_{s}}\right)^{2}}\right].
\label{eq:BalanceEq_tensor}
\end{equation}

The corresponding frequency-domain integral that defines the phase correction is then
\begin{equation}
I\left(  f\right) =\int_{\infty}^{\pi f}\frac{\left(  2\pi f-2\omega
_{s}\right)}{\omega_{s}^{\frac{11}{3}}\left[ 1-  \sqrt{1-\left(  \frac{m_{\Psi}c}{2\omega_{s}}\right)  ^{2}}\right]}  d\omega_{s}. \label{eq:integral_psi}
\end{equation}
Defining $x=\left(  \pi f\right)  ^{2},\text{ }\bar{m}=\left(  \frac{m_{\Psi}c}%
{2}\right)  ^{2}$ and expanding the solution of (\ref{eq:integral_psi}) around \(x_{0}=a\bar{m}\) with \(a>1\), we get
\begin{equation}
I(x)=\frac{3x^{-\frac{5}{6}}}{20}
\!\left[\bar{C}_{0}+\bar{C}_{1}\!\left(\frac{x}{a\bar{m}}-1\right)+\ldots\right], \label{eq:solution_psi_integral}
\end{equation}

where the constants \(\bar{C}_{0}\) and \(\bar{C}_{1}\) are defined as
\begin{align}
\bar{C}_{0}& =3+8F_{2}^{\text{ \ }1}\left( -\frac{1}{2},\frac{5}{6},\frac{11%
}{6},\frac{1}{a}\right)   \label{eq:C_0_bar} \\
& -5F_{2}^{\text{ \ }1}\left( -\frac{1}{2},\frac{4}{3},\frac{7}{3},\frac{1}{a%
}\right) ,  \nonumber \\
\bar{C}_{1}& =\frac{10}{77a}\left[ 14F_{2}^{\text{ \ }1}\left( \frac{1}{2},%
\frac{11}{6},\frac{17}{6},\frac{1}{a}\right) \right.   \label{eq:C_1_bar} \\
& \left. -11F_{2}^{\text{ \ }1}\left( \frac{1}{2},\frac{7}{3},\frac{10}{3},%
\frac{1}{a}\right) \right] .  \nonumber
\end{align}
Substituting these coefficients into Eq.~(\ref{eq:solution_psi_integral}), the frequency-domain phase becomes
\begin{align}
\Sigma_{+}&=2\pi f\!\left(t_{c}+\frac{r}{c}\right)-2\Phi_{c}-\frac{\pi}{4}
+\frac{3}{128}\!\left(\frac{G\mathcal{M}_{c}\pi f}{c^{3}}\right)^{-\frac{5}{3}}\!\times\nonumber\\
&\left[\frac{1}{3}\bar{C}_{0}
+\frac{1}{3}\bar{C}_{1}\!\left(\frac{(2\pi f)^{2}}{a m_{\Psi}^{2}c^{4}}-1\right)+\ldots\right],
\label{eq:phase_tensor_complete_rewrite}
\end{align}
and the corresponding amplitude correction reads
\begin{equation}
A_{c}\approx
1+\frac{1}{a^{\frac{3}{2}}\sqrt{a-1}}
\left[\frac{a(2a-3)}{4}+\frac{(\pi f)^{2}}{(m_{\Psi}c)^{2}}\right].
\label{eq:amplitude_tensor_complete_rewrite}
\end{equation}

\section{Parametrized post-Einsteinian parameters and observational
constraints} \label{sec:PPE}

The ppE framework is a waveform model proposed by Yunes and Pretorius
\cite{Yunes:2009ke} that describes the gravitational waves emitted by a
binary system on a quasicircular orbit in metric theories of gravity. In the
original ppE framework, all deviations from GR in gravitational waveforms can
be expressed through a set of four post-Einstein parameters $\left(
\alpha\text{, }\beta\text{, }a\text{, }b\right)  $ whose waveform would be
given by%
\begin{equation}
h\left(  f\right)  =h_{GR}\left(  f\right)  \left[  1+\sum_j\alpha_j\left(
\frac{G\mathcal{M}_{c}\pi f}{c^{3}}\right)  ^{\frac{a_j}{3}}\right]  e^{i\beta_j\left(
\frac{G\mathcal{M}_{c}\pi f}{c^{3}}\right)  ^{\frac{b_j}{3}}},\label{eq: PPE_original}%
\end{equation}
where $h_{GR}\left(  f\right)  $ is the GR waveform. This framework adjusts any given GR waveform model by introducing two classes of theoretical parameters: exponent parameters (a, b), which define the specific type of deviation from GR, and amplitude parameters ($\alpha$, $\beta$), which regulate the magnitude of the GR deformation. Moreover, it is worth
noting that the original ppE framework includes only the two tensor polarizations $h_{+}$ and $h_{\times}$. However,
extended versions of this framework were developed in Refs.
\cite{Bonilla:2022dyt, Loutrel:2022xok,
Maggio:2022hre, Mehta:2022pcn, Mezzasoma:2022pjb}.

\subsubsection{GR + massive spin-0 mode}
From Eq. (\ref{eq:phase_scalar_complete}) and Eq. (\ref{eq:amplitude_scalar_complete}), we obtain four distinct sets of ppE parameters corresponding to the scalar contribution in quadratic gravity.
\begin{align}
\alpha_1&=\frac{a(2a-3)}{72a^{\frac{3}{2}}\sqrt{a-1}}, \ \ a_1 = 0\\
\alpha_2 &= \frac{1}{18a^{\frac{3}{2}}\sqrt{a-1}}\left(\frac{c^2}{m_{\Phi}G\mathcal{M}_{c}}\right)^2, \ \ a_2 = 6 \\
\beta_{1} &= \frac{3}{128} \left( \frac{C_{0} - C_{1}}{54} - 1 \right), 
\quad b_{1} = -5, \\
\beta_{2} &= \frac{C_1}{576a} 
\left( 
\frac{c^{2}}{m_{\Phi}G \mathcal{M}_{c}}
\right)^{2},
\ \ b_{2} = 1,
\end{align}
where ${C}_{0}$ and ${C}_{1}$ are given by Eqs. (\ref{eq:C_0}) and (\ref{eq:C_1}). The values of \( b \) indicate deviations from General Relativity at the 0PN order (\( b = -5 \)) and at the 3PN order (\( b = 1 \)).

To obtain observational constraints on the deformation parameters from gravitational wave (GW) data, we need to relate the phase deformation in ppE framework to the gIMR\footnote{The gIMR (generalized IMRPhenom) waveform model is a phenomenological extension of the IMRPhenom formalism used by the LVK Collaboration. It introduces parametrized deviations from general relativity in the inspiral–merger–ringdown phases, allowing for consistency tests of GR with GW data.} waveform parameters. As described in Ref.~\cite{Yunes:2016jcc}, this relation is given by
\begin{equation}
\beta = \frac{3}{128\,\eta} \, \varphi_n \, \delta\varphi_n,
\end{equation}
with
\begin{equation}
b = n - 5,
\end{equation}
where \(\eta = \frac{m_1 m_2}{(m_1 + m_2)^2}\) is the symmetric mass ratio, \(\varphi_n\) are PN coefficients that depend on the component masses and spins, and \(\delta \varphi_n\) represents the fractional deviation from the GR PN coefficient.

Therefore, considering the phase variation due to the 3PN term, we obtain
\begin{equation}
\beta_2 = \frac{3}{128\,\eta} \, \varphi_6(\eta) \, \delta \varphi_6.
\label{eq:beta6_phi6}
\end{equation}

As described in Ref.~\cite{Arun:2004hn}, the coefficient \(\varphi_6(\eta)\) is given by
\begin{align}
\varphi_6(\eta) &= 
\frac{11583231236531}{4694215680}
- \frac{640}{3}\pi^{2}
- \frac{6848}{21}\gamma_E \\
&+ \left(-\frac{15737765635}{3048192} + \frac{2255}{12}\pi^{2}\right)\eta \nonumber
\\
&+ \frac{76055}{1728}\eta^{2}
- \frac{127825}{1296}\eta^{3} \nonumber,
\end{align}
where \(\gamma_E = 0.57721566\) is the Euler–Mascheroni constant. Thus, if we have an upper bound $|\delta \varphi_6| < k$, where $k$ corresponds to the 90\% marginalized constraint on the agnostic ppE parameter, we can establish corresponding constraints on the parameters of the theory. Since we aim to find order of magnitude constraints, we fix the mass parameters at their modes and consider the constraint on $k$ alone.

Therefore, in this case, the constraint can be expressed as
\begin{equation}
m_{\Phi}^{2} > \frac{2}{27} \, \frac{\eta}{\varphi_{6}(\eta)} \, \frac{1}{k} \,
\frac{C_{1}(a)}{a} \left( \frac{c^{2}}{G \mathcal{M}_{c}} \right)^{2}.
\end{equation}
To derive a constraint on the scalar field mass, and consequently on the parameter $\gamma$, it is necessary to fix the value of $a$, in addition to selecting a specific gravitational-wave event to determine the parameters $\eta$ and $k$. Since the massive field contributions are treated as small perturbative corrections to General Relativity, values of \( a \) close to unity are expected to provide better approximations. Nevertheless, we consider different values of \( a \) to derive the corresponding constraints and to illustrate this behavior more clearly.

For the observational input, we consider two gravitational-wave events: GW170817~\cite{LIGOScientific:2018dkp} and GW250114~\cite{LIGOScientific:2025obp}. The former is selected because lower chirp masses typically yield tighter constraints on the parameters of interest\footnote{In addition to the chirp mass, smaller values of parameter $k$ and values of the coupling parameter $a$ close to unity also contribute to improving the bounds on the theoretical parameters.}, while the latter corresponds to the event with the highest signal-to-noise ratio (SNR) observed to date, thus providing a complementary bound.

For the GW170817 event, using $\mathcal{M}_c = 1.18\,M_{\odot}$, $\eta = 0.24879$, and $k = 2$, we derive the constraints listed in Table~\ref{tab:constraints1}. The value of $k$ is taken from the 3PN correction reported in Fig.~2 of Ref.~\cite{LIGOScientific:2018dkp}. Similarly, for the GW250114 event, adopting $\mathcal{M}_c = 28.6\,M_{\odot}$, $\eta = 0.2499$, and $k = 0.08$, the resulting constraints are given in Table~\ref{tab:constraints2}, with the corresponding $k$ extracted from the 3PN correction (dark-blue star) in Fig.~5 of Ref.~\cite{LIGOScientific:2025obp}.

\begin{table}[h!]
\centering
\begin{tabular}{|c|c|c|}
\hline
\textbf{\( a \)} & \textbf{\( m_{\Phi} \, (\gtrsim) \, [\mathrm{m^{-1}}] \)} & \textbf{\( \gamma \, (\lesssim) \, [\mathrm{m^{2}}] \)} \\ 
\hline
1.01 & \( 2. \times 10^{-6} \) & \( 1. \times 10^{11} \) \\ 
\hline
1.50 & \( 1. \times 10^{-6} \) & \( 3. \times 10^{11} \) \\ 
\hline
2.00 & \( 7. \times 10^{-7} \) & \( 6. \times 10^{11} \) \\ 
\hline
\end{tabular}
\caption{Constraints on the scalar field mass \( m_{\Phi} \) and the coupling parameter \( \gamma \) for the GW170817 event.}
\label{tab:constraints1}
\end{table}

\begin{table}[h!]
\centering
\begin{tabular}{|c|c|c|}
\hline
\textbf{\( a \)} & \textbf{\( m_{\Phi} \, (\gtrsim) \, [\mathrm{m^{-1}}] \)} & \textbf{\( \gamma \, (\lesssim) \, [\mathrm{m^{2}}] \)} \\ 
\hline
1.01 & \( 3. \times 10^{-7} \) & \( 2. \times 10^{12} \) \\ 
\hline
1.50 & \( 2. \times 10^{-7} \) & \( 8. \times 10^{12} \) \\ 
\hline
2.00 & \( 1. \times 10^{-7} \) & \( 1. \times 10^{13} \) \\ 
\hline
\end{tabular}
\caption{Constraints on the scalar field mass \( m_{\Phi} \) and the coupling parameter \( \gamma \) for the GW250114 event.}
\label{tab:constraints2}
\end{table}

The results obtained in this work can be directly compared with those derived on astrophysical scales reported in the literature.  
In Refs.~\cite{Naf:2011za, Berry:2011pb, Vilhena:2021bsx}, constraints of $\gamma \lesssim 10^{17}\,\mathrm{m}^{2}$ and $\gamma \lesssim 10^{16}\,\mathrm{m}^{2}$ were established, respectively. In contrast, our constraints reach the order of \(10^{11}\)–\(10^{13}\,\mathrm{m^{2}}\), representing an improvement of several orders of magnitude.

The stronger constraints obtained in this work, compared to those reported in earlier studies, mainly result from the different physical regimes being examined. The scalar-mode limits in Refs.~\cite{Naf:2011za, Vilhena:2021bsx} come from pulsar binaries such as the Hulse--Taylor system, where the inspiral occurs at a very early stage, millions of years before coalescence, with typical orbital separations of about $10^{9}\,\mathrm{m}$. Similarly, the constraint reported in Ref.~\cite{Berry:2011pb} is derived from deviations in the perihelion precession of Mercury, a system characterized by a typical Mercury--Sun separation of order $10^{10}\,\mathrm{m}$. In contrast, the gravitational-wave events analyzed here probe systems much closer to merger, where the relevant length scales are several orders of magnitude smaller, only a few kilometers or less. At these smaller separations, additional degrees of freedom are more strongly excited, naturally leading to tighter constraints.

In Ref.~\cite{Abac:2025saz}, upper bounds on the beyond-GR parameters $\delta{\varphi}_n$ were derived during the inspiral phase of the GW150914 event, covering the range from $-1$PN to $3.5$PN order.  
In that study, the authors considered the projected sensitivity of the third-generation detector Einstein Telescope (ET), a proposed underground observatory featuring a triangular configuration of three $10$~km interferometers arranged at $60^\circ$ angles.  
This design provides improved sky coverage, enhanced polarization sensitivity, and a substantially higher signal-to-noise ratio compared to current detectors.  

For GW150914~\cite{LIGOScientific:2016lio}, it was shown that the ET improves over the Advanced LIGO’s first observing run by approximately two orders of magnitude for all testing coefficients.  
Following this improvement, and adopting the same approach for the GW170817 event, we find that for the scenario considered here the deformation parameter satisfies $\delta{\varphi}_6 < 0.02$.  
This allows us to project the corresponding bounds on the theory parameters $m_\Phi$ and $\gamma$, which are summarized as
\begin{table}[h!]
\centering
\begin{tabular}{|c|c|c|}
\hline
\textbf{\( a \)} & \textbf{\( m_{\Phi} \, (\gtrsim) \, [\mathrm{m^{-1}}] \)} & \textbf{\( \gamma \, (\lesssim) \, [\mathrm{m^{2}}] \)} \\ 
\hline
1.01 & \( 2. \times 10^{-5} \) & \( 1. \times 10^{9} \) \\ 
\hline
1.50 & \( 1. \times 10^{-5} \) & \( 3. \times 10^{9} \) \\ 
\hline
2.00 & \( 7. \times 10^{-6} \) & \( 6. \times 10^{9} \) \\ 
\hline
\end{tabular}
\caption{Projected constraints on the scalar field mass \( m_{\Phi} \) and coupling parameter \( \gamma \), assuming Einstein Telescope sensitivity for GW170817.}
\label{tab:project_constraints1}
\end{table}

\subsubsection{GR + massive spin-2 mode}

For the tensorial massive field, the procedure is entirely analogous to that of the scalar case, differing mainly in the sign and magnitude of the corrections to the waveform. From the phase and amplitude expressions, we obtain the following ppE parameters:
\begin{align}
\alpha_1 &= \frac{a(2a-3)}{4a^{3/2}\sqrt{a-1}}, \quad a_1 = 0,\\
\alpha_2 &= \frac{1}{a^{3/2}\sqrt{a-1}}\!\left(\frac{c^{2}}{m_{\Psi}G\mathcal{M}_{c}}\right)^{2}, \quad a_2 = 6,\\
\beta_{1} &= \frac{3}{128}\!\left(\frac{\bar{C}_{0}-\bar{C}_{1}}{3}-1\right), \quad b_{1}=-5,\\
\beta_{2} &= \frac{\bar{C}_{1}}{32a}\!
\left(\frac{c^{2}}{m_{\Psi}G\mathcal{M}_{c}}\right)^{2}, \quad b_{2}=1,
\end{align}
where $\bar{C}_{0}$ and $\bar{C}_{1}$ are defined in Eqs.~(\ref{eq:C_0_bar}) and (\ref{eq:C_1_bar}).  
As in the scalar case, the parameters \(b=-5\) and \(b=1\) correspond, respectively, to deviations at the 0PN and 3PN orders.

The corresponding constraint on the tensor field mass is given by
\begin{equation}
m_{\Psi}^{2} > \frac{4}{3}\,\frac{\eta}{\varphi_{6}(\eta)}\,\frac{1}{k}\,
\frac{\bar{C}_{1}(a)}{a}
\left(\frac{c^{2}}{G\mathcal{M}_{c}}\right)^{2}.
\label{eq:mPsi_constraint}
\end{equation}

Using the same events and parameters as in the scalar analysis, namely GW170817 ($\mathcal{M}_{c}=1.18\,M_{\odot}$, $\eta=0.24879$, $k=2$) and GW250114 ($\mathcal{M}_{c}=28.6\,M_{\odot}$, $\eta=0.2499$, $k=0.08$), we obtain the constraints summarized in Tables~\ref{tab:constraints3} and \ref{tab:constraints4}.

\begin{table}[h!]
\centering
\begin{tabular}{|c|c|c|}
\hline
\textbf{\( a \)} & \textbf{\( m_{\Psi} \, (\gtrsim) \, [\mathrm{m^{-1}}] \)} & \textbf{\( \alpha \, (\lesssim) \, [\mathrm{m^{2}}] \)} \\ 
\hline
1.01 & \( 7. \times 10^{-6} \) & \( 2. \times 10^{10} \) \\ 
\hline
1.50 & \( 4. \times 10^{-6} \) & \( 6. \times 10^{10} \) \\ 
\hline
2.00 & \( 3. \times 10^{-6} \) & \( 1. \times 10^{11} \) \\ 
\hline
\end{tabular}
\caption{Constraints on the tensor field mass \( m_{\Psi} \) and the coupling parameter \( \alpha \) for the GW170817 event.}
\label{tab:constraints3}
\end{table}

\begin{table}[h!]
\centering
\begin{tabular}{|c|c|c|}
\hline
\textbf{\( a \)} & \textbf{\( m_{\Psi} \, (\gtrsim) \, [\mathrm{m^{-1}}] \)} & \textbf{\( \alpha \, (\lesssim) \, [\mathrm{m^{2}}] \)} \\ 
\hline
1.01 & \( 1. \times 10^{-6} \) & \( 5. \times 10^{11} \) \\ 
\hline
1.50 & \( 9. \times 10^{-7} \) & \( 1. \times 10^{12} \) \\ 
\hline
2.00 & \( 6. \times 10^{-7} \) & \( 3. \times 10^{12} \) \\ 
\hline
\end{tabular}
\caption{Constraints on the tensor field mass \( m_{\Psi} \) and the coupling parameter \( \alpha \) for the GW250114 event.}
\label{tab:constraints4}
\end{table}

The obtained results improve upon those presented in Ref.~\cite{Alves:2022yea}, where two independent methods yielded $\alpha \lesssim 1.1\times10^{21}\,\mathrm{m^{2}}$ (waveform analysis) and $\alpha \lesssim 1.1\times10^{13}\,\mathrm{m^{2}}$ (coalescence-time method).  
In contrast, our constraints reach the order of \(10^{10}\)–\(10^{12}\,\mathrm{m^{2}}\), thus representing an improvement of several orders of magnitude. 

Just as in the scalar case, the tensor-mode constraints also show differences in magnitude when compared with previous methods. In this case, Ref.~\cite{Alves:2022yea} provides two types of bounds: one obtained from direct waveform comparison (WF) and another from the time-to-coalescence (TC). The WF bound ($\alpha < 10^{21}\,\mathrm{m^2}$) is relatively weak because the extra tensor mode that contributes to the signal is generated in a region that is still far from merger. The TC bound is much stronger ($\alpha < 10^{13}\,\mathrm{m^2}$), and it is close to the value we obtain using the ppE approach ($\alpha < 10^{11}\,\mathrm{m^2}$). Both methods probe near-merger length scales, but the TC estimate relies on a coarse approximation, whereas the 3PN variation used in the ppE framework provides a more accurate measurement. This explains why our constraints are significantly stronger than those reported in earlier studies\footnote{It is important to emphasize that additional factors, such as the chirp mass and the signal-to-noise ratio, also influence the strength of the constraints in both cases.
}.

Finally, considering the projected sensitivity of the Einstein Telescope (ET), we can forecast the corresponding bounds on the tensorial field.  
Assuming the same improvement factor as in the previous case, the projected bounds are summarized in Table~\ref{tab:project_constraints_tensor}.

\begin{table}[h!]
\centering
\begin{tabular}{|c|c|c|}
\hline
\textbf{\( a \)} & \textbf{\( m_{\Psi} \, (\gtrsim) \, [\mathrm{m^{-1}}] \)} & \textbf{\( \alpha \, (\lesssim) \, [\mathrm{m^{2}}] \)} \\ 
\hline
1.01 & \( 7. \times 10^{-5} \) & \( 2. \times 10^{8} \) \\ 
\hline
1.50 & \( 4. \times 10^{-5} \) & \( 6. \times 10^{8} \) \\ 
\hline
2.00 & \( 3. \times 10^{-5} \) & \( 1. \times 10^{9} \) \\ 
\hline
\end{tabular}
\caption{Projected constraints on the tensor field mass \( m_{\Psi} \) and coupling parameter \( \alpha \), assuming Einstein Telescope sensitivity for GW170817.}
\label{tab:project_constraints_tensor}
\end{table}

\section{Conclusion} \label{sec:conclusion}
In this work, we studied the gravitational waveforms emitted during the inspiral phase of compact binary systems in non-relativistic quasi-circular orbits within quadratic gravity. Our analysis is based on the assumption that the massive modes are small perturbations to General Relativity (GR), implying that their direct effect on the waveform is negligible, but they influence the orbital dynamics, which in turn modifies the waveform observed in the massless field. Additionally, the construction is performed within the quadrupole approximation of quadratic gravity, so the dominant contribution arises from the quadrupole term while higher-order corrections are subdominant. We also assume that the vacuum solution is Schwarzschild, although in principle quadratic gravity admits alternative vacuum solutions that are not considered here.

Within this framework, we mapped the results into the parameterized post-Einsteinian (ppE) framework, which allowed us to extract new observational constraints on the theory parameters. We identified three distinct polarizations: the combinations $\{h_+,\Psi_+\}$ and $\{h_\times,\Psi_\times\}$ arising from the tensor fields, and the scalar mode $\Phi$ from the massive scalar field. Our analysis is restricted to transverse tensor modes due to the unphysical nature of longitudinal components (see Ref.~\cite{Alves:2024gwi} for further discussion), leaving the theory with three physical degrees of freedom. The massive fields exhibit two regimes: an oscillatory regime responsible for gravitational wave emission and a damped regime characterized by exponential decay. Using the associated energy balance equation, we quantified the energy loss due to gravitational wave emission.

Employing the stationary phase approximation (SPA), we computed the Fourier transform of the massless tensor polarizations and derived analytical expressions for their phases, $\Sigma_+$ and $\Sigma_\times$. When scalar and massive tensorial modes are treated as perturbations to GR, the massless tensorial polarizations can be mapped onto the ppE framework, yielding the parameters $(a, b, \alpha, \beta)$. In Sec.~\ref{sec:PPE}, we explicitly show that the applicability of the ppE framework depends on the parameters of the theory. Our results also extend those of Ref.~\cite{Liu:2020moh}, where the mapping of massive Brans-Dicke theory to the ppE framework was limited to scalar fields with sufficiently small masses.

After mapping the quadratic gravity theory onto the ppE framework, we derived observational constraints on the theory parameters using the GW170817 and GW250114 events.  
For scalar and tensor massive fields, the most stringent bounds were obtained from GW170817, yielding 
\(\gamma \lesssim 1.26 \times 10^{11}\,\mathrm{m^{2}}\) and 
\(\alpha \lesssim 2.10 \times 10^{10}\,\mathrm{m^{2}}\), respectively (Tables~\ref{tab:constraints1} and \ref{tab:constraints3}).  
These limits improve upon previous astrophysical-scale analyses by several orders of magnitude.  
Constraints from the GW250114 event provide complementary bounds, with 
\(\gamma \lesssim 2.97 \times 10^{12}\,\mathrm{m^{2}}\) and 
\(\alpha \lesssim 4.94 \times 10^{11}\,\mathrm{m^{2}}\) (Tables~\ref{tab:constraints2} and \ref{tab:constraints4}).  
Finally, the projected constraints based on the enhanced sensitivity of the Einstein Telescope (ET) yield 
\(\gamma \lesssim 1.26 \times 10^{9}\,\mathrm{m^{2}}\) and 
\(\alpha \lesssim 2.10 \times 10^{8}\,\mathrm{m^{2}}\) (Tables~\ref{tab:project_constraints1} and \ref{tab:project_constraints_tensor}), 
representing the most stringent forecasted bounds within our analysis.

\section*{Acknowledgements}

MFSA thanks \textit{Fundação de Amparo à Pesquisa e Inovação do Espírito Santo} (FAPES, Brazil) for support. LGM thanks \textit{Conselho Nacional de Desenvolvimento Científico e Tecnológico} (CNPq, Brazil) for partial financial support—Grant: 307901/2022-0 (LGM). DCR thanks \textit{Centro Brasileiro de Pesquisas Físicas} (CBPF) and \textit{Núcleo de Informação C\&T e Biblioteca} (NIB/CBPF) for hospitality, where part of this work was done. He also acknowledges CNPq (Brazil), FAPES (Brazil) and \textit{Fundação de Apoio ao Desenvolvimento da Computação Científica} (FACC, Brazil) for partial support.

\bibliographystyle{apsrev4-1}
\bibliography{main}

\end{document}